# Polyp-SAM: Transfer SAM for Polyp Segmentation

Yuheng Li[a], Mingzhe Hu[b] and Xiaofeng Yang[a,b,c*]
[a]Department of Biomedical Engineering, Emory University, GA, Atlanta, USA
[b]Department of Computer Science and Informatics, Emory University, GA, Atlanta, USA
[c]Department of Radiation Oncology, Winship Cancer Institute, School of Medicine, Emory University, GA, Atlanta, USA
*Email: xiaofeng.yang@emory.edu

## Abstract

Colon polyps are considered important precursors for colorectal cancer. Automatic segmentation of colon polyps can significantly reduce the misdiagnosis of colon cancer and improve physician annotation efficiency. While many methods have been proposed for polyp segmentation, training large-scale segmentation networks with limited colonoscopy data remains a challenge. Recently, the Segment Anything Model (SAM) has recently gained much attention in both natural and medical image segmentation. SAM demonstrates superior performance in several image benchmarks and therefore shows great potential for medical image segmentation. In this study, we propose Poly-SAM, a finetuned SAM model for polyp segmentation, and compare its performance to several state-of-the-art polyp segmentation models. We also compare two transfer learning strategies of SAM with and without finetuning its encoders. Evaluated on five public datasets, our Polyp-SAM achieves state-of-the-art performance on two datasets and impressive performance on three datasets, with dice scores all above 88%. This study demonstrates the great potential of adapting SAM to medical image segmentation tasks. We plan to release the code and model weights for this paper at: https://github.com/ricklisz/Polyp-SAM.

## 1. Introduction

Colorectal cancer (CRC) is a common gastrointestinal malignancy with high morbidity and a low 5-year survival rate if diagnosed late. CRC is the second leading cause of cancer-related deaths in United States with more than 106,180 cases and 52,580 deaths in 2022 [1]. The presence of Colon polyp is an important factor in causing CRC. Early detection and removal of CRC can significantly reduce the mortality rate and improve patient outcomes [2]. Colonoscopy is considered the gold standard for CRC diagnosis. It can identify and eliminate colorectal polyps, which can prevent further damage to the surrounding tissues and lower the chance of CRC. Currently, the screening of colorectal polyps relies on manual inspection of endoscopic images. However, this method is tedious and time-consuming, and subject to clinician's inter-reader variability.

Computer-aided diagnosis (CAD) for colonoscopy can effectively improve annotation efficiency and reduce time-to-diagnosis [3]. Recent advances in deep learning models have resulted in many neural networks for medical image segmentation. One of the most popular architectures is U-Net [4], a convolutional neural network with an encoder path and a decoder path. The encoder path captures the context of the image producing feature maps and it is made up of a stack of convolution and max pooling layers. The decoder path enables precise localization using transposed convolutions. Zhou et al. presented U-Net++ [5], which is a nested architecture for medical image segmentation. However, annotation time and costs remain significant obstacles for collecting and curating medical datasets to train large-scale segmentation models, primarily because trained physicians must typically provide careful mask annotations for images.

Transfer learning leverages the knowledge learned from large-scale natural image datasets to solve specific problems in the medical domain [6, 7]. Recent advances in large foundational language models such as Chat-GPT and GPT-4 have shown impressive performances in various language tasks [8]. With their superior transferability to different tasks, many have explored transferring foundational language models to medical tasks [9, 10]. Segment Anything Model (SAM) is proposed as a foundation model for image segmentation that generates high-quality object masks from input prompts (e.g., points, boxes, and masks) [11]. Due to its promising performance in several computer vision benchmarks, SAM has attracted much attention in medical image segmentation [12-14]. In this paper, we propose Poly-SAM, a foundational vision model for polyp segmentation using transfer learning. We finetune SAM on a collection of multi-center colonoscopy images. We show that with our finetuned SAM can outperform existing methods for polyp segmentation and demonstrate superior generalizability to multi-institutional data.

## 2. Method

### 2.1 Datasets

To validate the effectiveness of our finetuning method, we conducted comparison experiments on five benchmark colonoscopy datasets: 1) Kvasir [15]: This dataset was collected by Vestre Viken Health Trust in Norway and consisting of 1,000 polyp images and their corresponding ground truth from colonoscopy video sequences. The ground truth was manually annotated by medical doctors and then verified by experienced gastroenterologists. The resolution of the images varies from 332 × 487 to

1920 × 1072 pixels. 2) CVC-ClinicDB [16]: This dataset contains 612 images collected from 29 colonoscopy video sequences with a resolution of 288×384. It was built in collaboration with the Hospital Clinic of Barcelona, Spain. 3) CVC-ColonDB [17]: This dataset consists of 380 polyp images and their corresponding ground truth with a resolution of 500×570. The images were generated from 15 different videos from which experts guaranteed that the extracted frames showed an obviously different viewpoint by rejecting similar frames and annotated the 380 frames manually. 4) ETIS [16] This dataset contains 196 polyp images with a size of 966 × 1225. 5). CVC-300 [17]: This dataset includes 60 polyp images with a resolution of 500×574. For fair comparison to previous studies, we followed the same data split settings in [18], using 900 images from Kvasir and 550 ones from CVC-ClinicDB to form the training set. The remaining images from the two datasets (i.e., Kvasir and CVC-ClinicDB) and the other three datasets (i.e., CVC-ColonDB, ETIS, and CVC-300) aere used for testing.

## 2.2. Segment Anything

Segment Anything Model (SAM) is a model that segments user-defined objects of interest with user prompts such as point, mask, bounding box, or text. SAM will also return a segmentation mask even if the prompt is missing or ambiguous. SAM has three components: an image encoder, a prompt encoder and a mask decoder. The image encoder of SAM uses a masked autoencoder (MAE) that has been modified to handle input images of up to 1024x1024 resolution. The ViT backbone consists of a series of self-attention layers to capture long-range dependencies between patches of the input and the the input image. This is followed by a set of feedforward layers that transform the output of the self-attention layers into a set of feature maps used by the mask decoder to generate segmentation masks. The prompt encoders are tailored for different user inputs such as points, boxes, texts, and masks. Each point is encoded by Fourier positional encoding and two learnable tokens for specifying foreground and background, respectively. The bounding box is encoded by the point encoding of its top-left corner and bottom-right corner. The free-form text is encoded by the pre-trained text-encoder in CLIP. The mask prompt has the same spatial resolution as the input image, which is encoded by convolution feature maps. Finally, the mask decoder employs a lightweight design, which consists of two transformer layers with a dynamic mask prediction head and an Intersection-over-Union (IoU) score regression head. The mask prediction head can generate three 4× downscaled masks, which correspond to the whole object, part, and subpart of the object.

SAM was trained using SA-1B dataset which was developed in three stages. First, a set of images was annotated by human annotators by clicking on objects and manually refining masks generated by SAM which at that point was trained using public datasets. Second, the annotators were asked to segment masks that were not confidently generated by SAM to increase the diversity of objects. The final set of masks was generated automatically by prompting the SAM model with a set of points distributed in a grid across the image and selecting confident and stable masks. SA-1B includes more than 1B masks from 11M licensed and privacy-preserving image. This dataset has 400× more masks than any existing segmentation dataset, resulting in a high potential for being able to segment objects of types that it has not seen during training, i.e., zero-shot learning ability.

## 2.3 Preprocessing and Prompt generation

Since the SAM encoder requires high-resolution input, each image was resized to 3×1024×1024 before training. The output masks from the SAM decoder were then resized to the original resolution of the image. Down-sampled masks were compared to the ground truth masks to calculate loss.

SAM relies on accurate prompts to generate high-quality masks. Previous investigations of SAM demonstrated that the bounding box is the most effective prompt for SAM. In our experiments, only the Kvasir dataset provided ground truth bounding boxes for each patient. For the remaining two datasets, we automatically extracted the bounding boxes from the ground truth masks. During training and evaluation, we directly utilized the bounding boxes to prompt SAM to generate polyp masks.

## 2.4 Polyp-SAM: Transfer SAM for polyp segmentation

To fully explore SAM's potential for medical image segmentation, we propose two strategies for transfer learning SAM (Fig. 1): 1) Pretraining only the mask decoder while freezing all encoders; 2) Pretraining the image encoder, prompt encoder, and mask decoder. For the base model, we evaluate both strategies using ViT-B. Next, we aim to finetune SAM's two base models (ViT-B and ViT-L) for polyp segmentation (Polyp-SAM-B and Polyp-SAM-L). During this step, we finetuned all components of SAM.

We trained using an effective batch size of 48 with gradient accumulation. The loss function used was a simple Dice Loss commonly used in segmentation tasks. We optimized using AdamW and used a learning rate scheduler with linear warm-up and cosine annealing. The learning rate was set at 4e-6. During training, we used only 80% of the data while the remaining 20% was used for validation to restore the best weights. The PC used to fine-tune the model has an x86_64 architecture with 12 CPUs, 6 cores per socket, and 2 sockets, for a total of 24 threads. The CPU model is an Intel Xeon E5-2603 v4 @ 1.70GHz with 30MB L3 cache. The system has a 46-bit physical and 48-bit virtual address size and supports virtualization via VT-x. The GPU is a Tesla V100-PCIE with 2 cards, each with a memory capacity of 32GB.

For evaluation, we used the Dice Similarity Coefficient (DSC) and mean Intersection over Union (mIoU) to evaluate the region overlap ratio between the ground truth and segmentation results, which are two commonly used segmentation metrics in polyp segmentation. mIoU measures the mean overlap between the predicted and ground truth masks by dividing the intersection of the masks by the union of the masks. DSC also measures the overlap between the predicted and ground truth masks, which is calculated by computing the harmonic mean of the precision and recall scores.

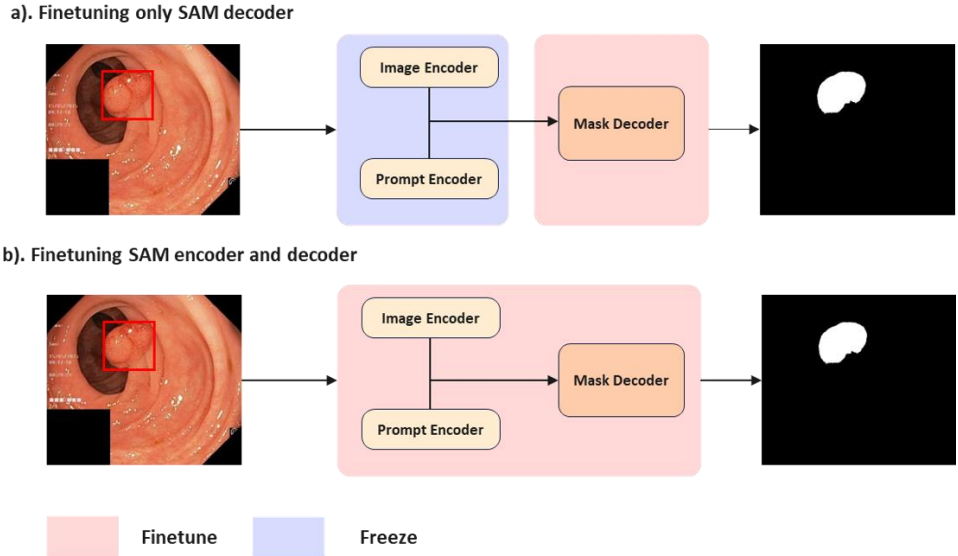

**Figure 1.** Our proposed transfer learning strategies for Poly-SAM. a). We freeze SAM's image encoder and prompt encoder and finetune only mask decoder. b). We finetune both encoders and decoder.

## 3. Result

### 3.1 Comparison of transfer learning strategies

We first compared the results of our two transfer learning strategies on Polyp-SAM-B using a multi-center dataset. As shown in Table 1, our model demonstrated better performance when all components were fine-tuned. However, we observed that the gains from fine-tuning all components were limited (0-3% in DSC and 0.1-4% in mIoU) compared to fine-tuning only the mask decoder. Therefore, it remains an open question whether SAM's encoders should be fine-tuned to achieve optimal performance.

**Table 1.** Comparison of transfer learning strategies for Polyp-SAM-B on multi-center data. Finetuning all components of SAM consistently leads to better performance metrics.

| Dataset | DSC | | mIoU | |
| --- | --- | --- | --- | --- |
| | Finetune Dec | Finetune Enc-Dec | Finetune Dec | Finetune Enc-Dec |
| Kvasir | 0.882 | 0.902 | 0.835 | 0.863 |
| CVC-ClinicDB | 0.890 | 0.921 | 0.837 | 0.877 |
| CVC-ColonDB | 0.891 | 0.894 | 0.835 | 0.843 |
| CVC-300 | 0.924 | 0.924 | 0.882 | 0.883 |
| ETIS | 0.885 | 0.903 | 0.834 | 0.852 |

## 3.2 Multi-center generalization of Polyp-SAM

We conducted two multi-center generalization studies of Polyp-SAM, where we compared two base models of SAM and fine-tuned all components. Our first study followed the data split strategy in [17] using mixed images from CVC-ColonDB, Kvasir-SEG, CVC-ClinicDB, CVC-300, and ETIS. The polyp segmentation metrics of all the compared algorithms on the test set are shown in Table 2. Our Polyp-SAM-B achieved state-of-the-art results in CVC-Colon DB (89.4% DSC), CVC-300 (92.4% DSC), and ETIS (90.3% DSC), outperforming all comparable methods. Polyp-SAM-B also achieved comparable results with other methods in CVC-ClinicalDB and Kvasir. Polyp-SAM-L achieved state-of-the-art results in CVC-300 with 92.9% DSC and 88.9% mIoU, and in ETIS with 90.5% DSC and 86.0% mIoU. Visualization of Polyp-SAM's segmentation results from mixed datasets are shown in Figure 2 and Figure 3. In general, we find that both Polyp-SAM-B and Polyp-SAM-L perform reasonably well across multi-center data. We also show some cases where Polyp-SAM fails to detect lesions in Figure 4. We suspect that Polyp-SAM still requires further finetuning in the case of multiple sporadic lesions.

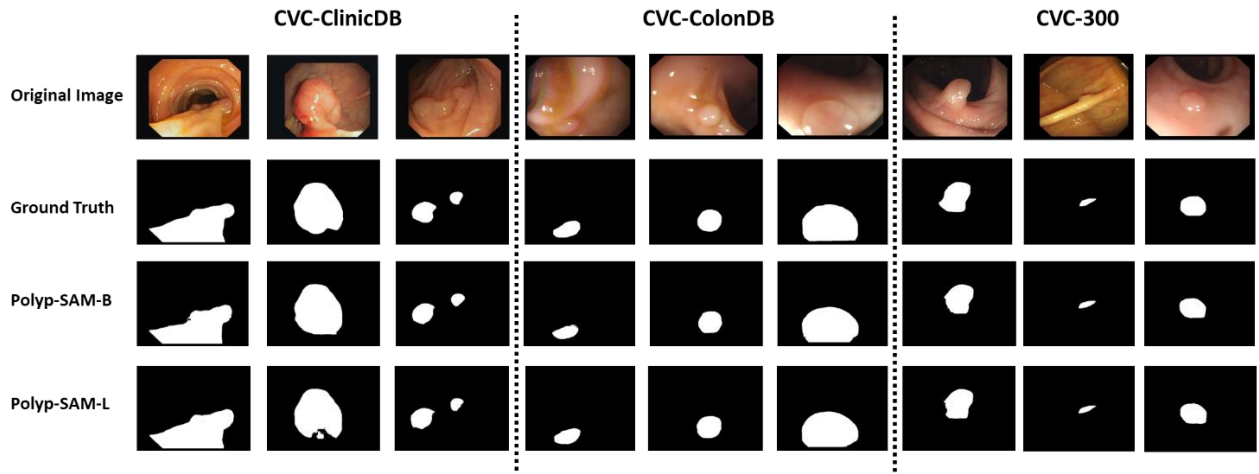

**Figure 2.** Segmentation results on CVC-ClincDB, CVC-ColonDB and CVC-300 for Polyp-SAM-B and Polyp-SAM-L.

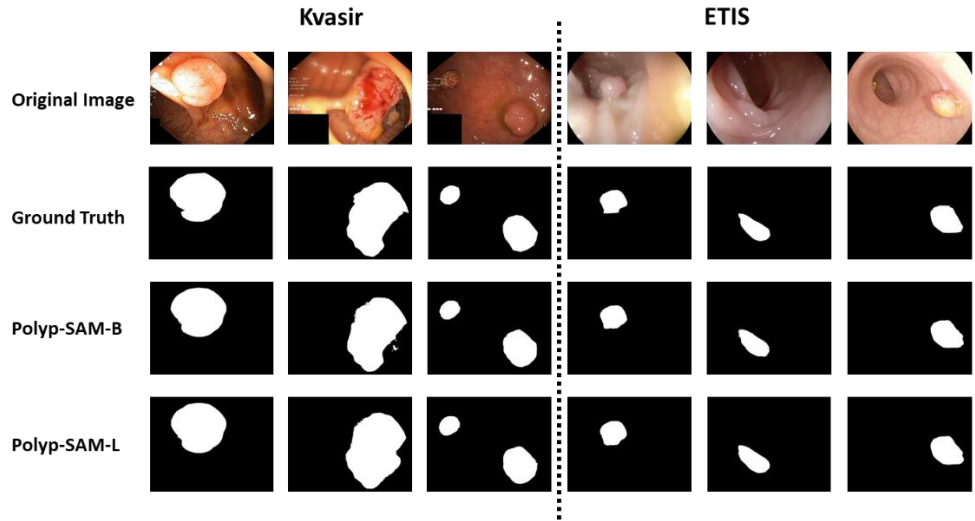

**Figure 3.** Segmentation results on Kvasir and ETIS for Polyp-SAM-B and Polyp-SAM-L.

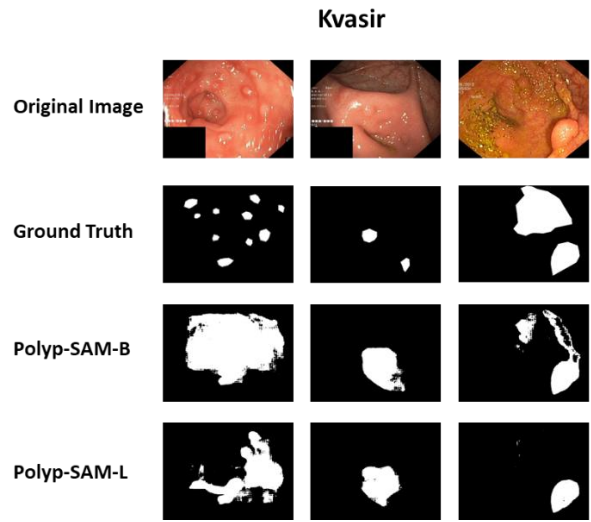

**Figure 4.** Failure segmentation results on Kvasir for Polyp-SAM-B and Polyp-SAM-L.

**Table 2.** Comparison of polyp segmentation results on the multi-center datasets (Kvasir, CVC-ClinicDB, CVC-ColonDB and CVC-300).

| Models | Kvasir | | CVC-ClinicDB | | CVC-ColonDB | | CVC-300 | | ETIS | |
|---|---|---|---|---|---|---|---|---|---|---|
| | DSC | mIoU | DSC | mIoU | DSC | mIoU | DSC | mIoU | DSC | mIoU |
| U-Net [4] | 0.818 | 0.746 | 0.823 | 0.755 | 0.504 | 0.436 | 0.710 | 0.627 | 0.398 | 0.335 |
| U-Net++ [19] | 0.821 | 0.744 | 0.794 | 0.729 | 0.482 | 0.408 | 0.707 | 0.624 | 0.401 | 0.344 |
| DCRNet [20] | 0.886 | 0.825 | 0.896 | 0.844 | 0.704 | 0.631 | 0.856 | 0.788 | 0.556 | 0.496 |
| C2FNet [21] | 0.886 | 0.831 | 0.919 | 0.872 | 0.724 | 0.650 | 0.874 | 0.801 | 0.699 | 0.624 |
| LDNet [22] | 0.887 | 0.821 | 0.881 | 0.825 | 0.740 | 0.652 | 0.869 | 0.793 | 0.645 | 0.551 |
| Polyp-PVT [23] | 0.917 | 0.864 | 0.948 | 0.905 | 0.808 | 0.727 | 0.900 | 0.833 | 0.787 | 0.706 |
| HSNet [24] | 0.926 | 0.877 | 0.937 | 0.887 | 0.810 | 0.735 | 0.903 | 0.839 | 0.808 | 0.734 |
| **Poly-SAM-B** | **0.902** | **0.863** | **0.921** | **0.877** | **0.894** | **0.843** | **0.924** | **0.882** | **0.903** | **0.852** |
| **Poly-SAM-L** | **0.901** | **0.864** | **0.917** | **0.874** | **0.889** | **0.838** | **0.929** | **0.889** | **0.905** | **0.860** |

For our second multi-center generalization study, we used cross-datasets, i.e. the original training set and test set respectively used images from different centers. We fine-tuned Polyp-SAM using Kvasir and CVC-ClinicDB datasets with 80% for training and 20% for validation. To evaluate Polyp-SAM's generalizability, we then evaluated it directly on CVC-ColonDB. The comparison of segmentation results is shown in Table 3. Our Polyp-SAM-B achieved state-of-the-art results when compared to other methods (90.6% DSC and 85.5% mIoU). Our Polyp-SAM-L achieved the second-best result (88.1% DSC and 82.5% mIoU). This result demonstrates Polyp-SAM's superior generalizability across different datasets. When using Colon-DB as the testing set, we noticed that Poly-SAM-B's performance did not increase when trained on more data (i.e., CVC-ClinicDB and CVC-300). Polyp-SAM-L gained a 0.8% increase in DSC and a 1.3% increase in mIoU. This result is expected since Polyp-SAM-L is more computationally complex and can benefit from a more diverse training dataset.

Table 3. Comparison of polyp segmentation results on the held-out CVC-ColonDB.

| Models | CVC-Colon DB | |
| --- | --- | --- |
| | DSC | mIoU |
| U-Net | 0.504 | 0.436 |
| U-Net++ | 0.482 | 0.408 |
| DCRNet | 0.704 | 0.631 |
| C2FNet | 0.724 | 0.650 |
| LDNet | 0.740 | 0.652 |
| Polyp-PVT | 0.808 | 0.727 |
| HSNet | 0.810 | 0.735 |
| **Poly-SAM-B** | **0.906** | **0.855** |
| **Poly-SAM-L** | **0.881** | **0.825** |

## 4. Discussion and conclusion

In this paper, we propose Polyp-SAM, a foundational model for polyp segmentation using colonoscopy images by finetuning SAM. Both Polyp-SAM-B and Polyp-SAM-L demonstrate excellent performance across five public colon polyp datasets when compared to existing methods. Although Polyp-SAM-L is much more computationally complex, we note that it did not outperform the lightweight Polyp-SAM-B in three out of five public datasets during finetuning. This result led us to believe that the lightweight ViT-B model of SAM is more suitable to be transferred to medical image applications. We also compare two strategies for transfer learning SAM (i.e., finetuning decoder only or finetuning all components of SAM). Our results indicated that while finetuning all components of SAM leads to better performance, the gains are still quite limited, and finetuning only the decoder achieves satisfactory performance.

However, our study has some limitations. As shown in Figure 4, Polyp-SAM still struggles with multiple sporadic lesions in colonoscopy images. We suspect that Polyp-SAM's performance can be further improved by adjusting mask decoders to output multiple segmentation masks. In addition, since SAM is based on learning visual patterns from user prompts, it becomes a major bottleneck for developing fully end-to-end polyp segmentation models using SAM. While our Polyp-SAMs require ground truth bounding boxes as prompts for optimal performance, we believe that our model still has clinical significance, as manual annotations are time-consuming and tedious for physicians to perform on each image. Given Polyp-SAM's state-of-the-art performance in five public datasets using cross-dataset evaluation, we believe that our model is robust to distribution shifts in various clinical colonoscopy data. Furthermore, SAM is an image segmentation model, while the clinical applications for colonoscopy should be video segmentation models. For SAM, combining multiple frames using spatial and temporal information remains to be investigated.